\pdfoutput=1
\documentclass[twocolumn,showkeys,showpacs,preprintnumbers,amsmath,amssymb,prl]{revtex4}
\usepackage{graphicx}
\usepackage{dcolumn}
\usepackage{bm}
\usepackage{amssymb}
\usepackage{amsthm}

\begin{document}

\title{Unusual magnetic behavior in ferrite hollow nanospheres}

\author{E. Lima Jr.\footnote{Presently at: Instituto de F\'{\i}sica, Universidade de S\~{a}o Paulo, Brazil}, J. M. Vargas, R. D. Zysler\footnote{corresponding author, e-mail: zysler@cab.cnea.gov.ar}}

\affiliation{Centro At\'{o}mico Bariloche and Instituto Balseiro,
8400 S. C. de Bariloche, RN, Argentina}

\author{H. R. Rechenberg}

\affiliation{Instituto de F\'{\i}sica, Universidade de S\~{a}o
Paulo, 05315-970, S\~{a}o Paulo, SP, Brazil.}

\author{J. Arbiol}

\affiliation{TEM-MAT Serveis Cientifico-t\'{e}cnicos, Universidad de
Barcelona, 08028, Barcelona, Spain.}

\author{G. F. Goya\footnote{corresponding author, e-mail: goya@unizar.es}, A. Ibarra, M. R. Ibarra}

\affiliation{Instituto de Nanociencia de Arag\'{o}n, Universidad de
Zaragoza, 50009, Zaragoza, Spain.}

\date{\today }

\begin{abstract}
We report unusual magnetic behavior in iron oxide hollow nanospheres
of 9.3 $nm$ in diameter. The large fraction of atoms existing at the
inner and outer surfaces gives rise to a high magnetic disorder. The
overall magnetic behavior can be explained considering the
coexistence of a soft superparamagnetic phase and a hard phase
corresponding to the highly frustrated cluster-glass like phase at
the surface regions.
\end{abstract}

\pacs{75.50.Tt, 75.50.Gg, 75.50.LK, 75.60.Ej}

\keywords{Fine Particles, Superparamagnetism, Magnetic ordering,
Magnetic Freezing}

\maketitle

Finite size-effect occurs in nanostructured materials as thin films,
nanoparticles and nanowires. The control of their morphology and
functionalities at the nanoscale is a prerequisite for some
biomedical applications that use nanoparticles as nanovectors for
drug delivery \cite{ARR07}. Spherical empty nanocapsules are
appealing for these applications because they could store larger
amounts of drug than solid NPs of the same size. The unique magnetic
phenomena reported for core-shell nanoparticles along the last years
have been usually assigned to the complex surface microstructure
and/or exchange interactions at the core/surface interface
\cite{MAR98,BAT02}. The magnetic behavior of the surface atoms is
characterized by the existence of broken symmetry and exchange bonds
which introduce structural and magnetic disorder and originate an
enhancement of the magnetic anisotropy and the coercive field
\cite{KOD99}. On the bases of the huge surface/bulk atomic ratio,
Hollow NanoSpheres (HNS) provide an excellent scenario to study the
competition between surface and bulk magnetism at nanoscale level
and open up new perspectives for theoretical developments. The
synthesis of HNS have been recently reported \cite{SUN07}, using
controlled oxidization of $Fe$ or $Fe_3O_4$ nanoparticles after
synthesis, at temperatures of 473-473 $K$. In this work we present a
different, low-temperature synthesis method for obtaining
monodisperse ferrite HNS without need of subsequent oxidization
process. The unusual magnetic behavior found in these particles can
be interpreted on the basis of soft and hard phases and the large
surface/bulk atomic ratio due to both inner and outer surfaces of a
hollow sphere.

\begin{figure}
\begin{center}
\includegraphics[bb=-20bp 0bp 670bp 610bp,
clip,width=1.0\columnwidth]{{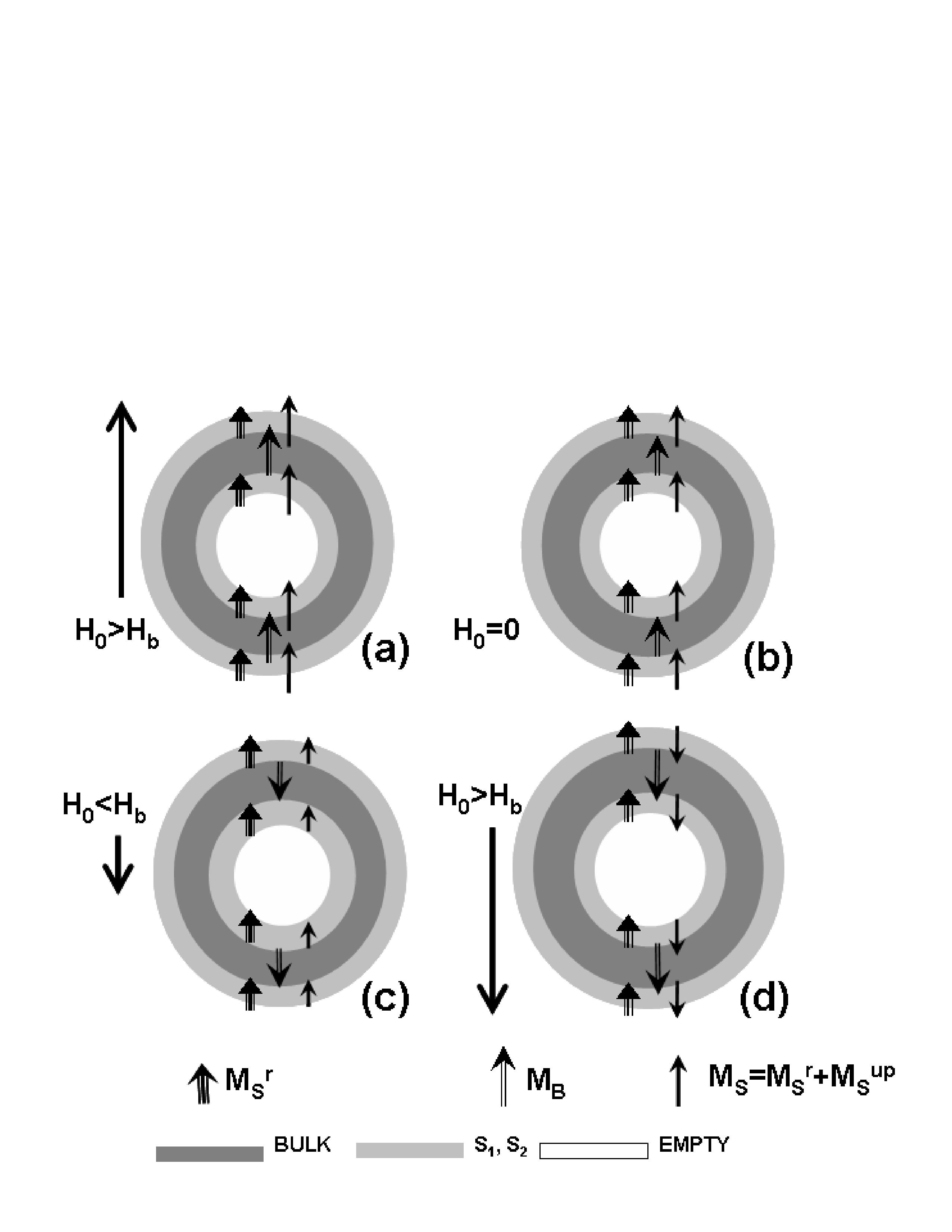}}
\caption{\label{FIG:DESENHO} Schematic representation of the
magnetization process of the sample after FC below $T_f$. (See
explanation in the text).}
\end{center}
\end{figure}

Ferrite HNS were prepared by modifications at the high-temperature
organic-phase synthesis from the precursor $Fe(acac)_3$ at
phenylether (boiling point $\sim$ 533-543 $K$) in the presence of a
long-chain alcohol (1,2-hexadecanediol) and oleic acid and
oleylamine surfactants \cite{SUN04}, using a molar ratio
precursor:surfactant of $1$:$9$  to control the final diameter of
the particles \cite{VAR05}. The synthesis lasted 30 minutes in argon
flux ($\sim$0.5 $L/min.$), but differently from those described in
the literature, a (noncontrolled) temperature reduction was induced
during the synthesis procedure. Final HNS were coated by surfactant
molecules, avoiding agglomeration and increasing the chemical
stability of the surface. They were further dispersed by dilution in
toluene and alcohol dilutes polyethylemine (PEI). After that the
solvents were left to evaporate, being stirred from time to time. At
the end, HNS dispersed to 5 $\%$ $wt$ in PEI were obtained; this
dilution is sufficient to ensure a negligible dipolar interaction.
\begin{figure}
\begin{center}
\includegraphics[bb=0bp 2bp 500bp 500bp,
clip,width=1.0\columnwidth,keepaspectratio]{{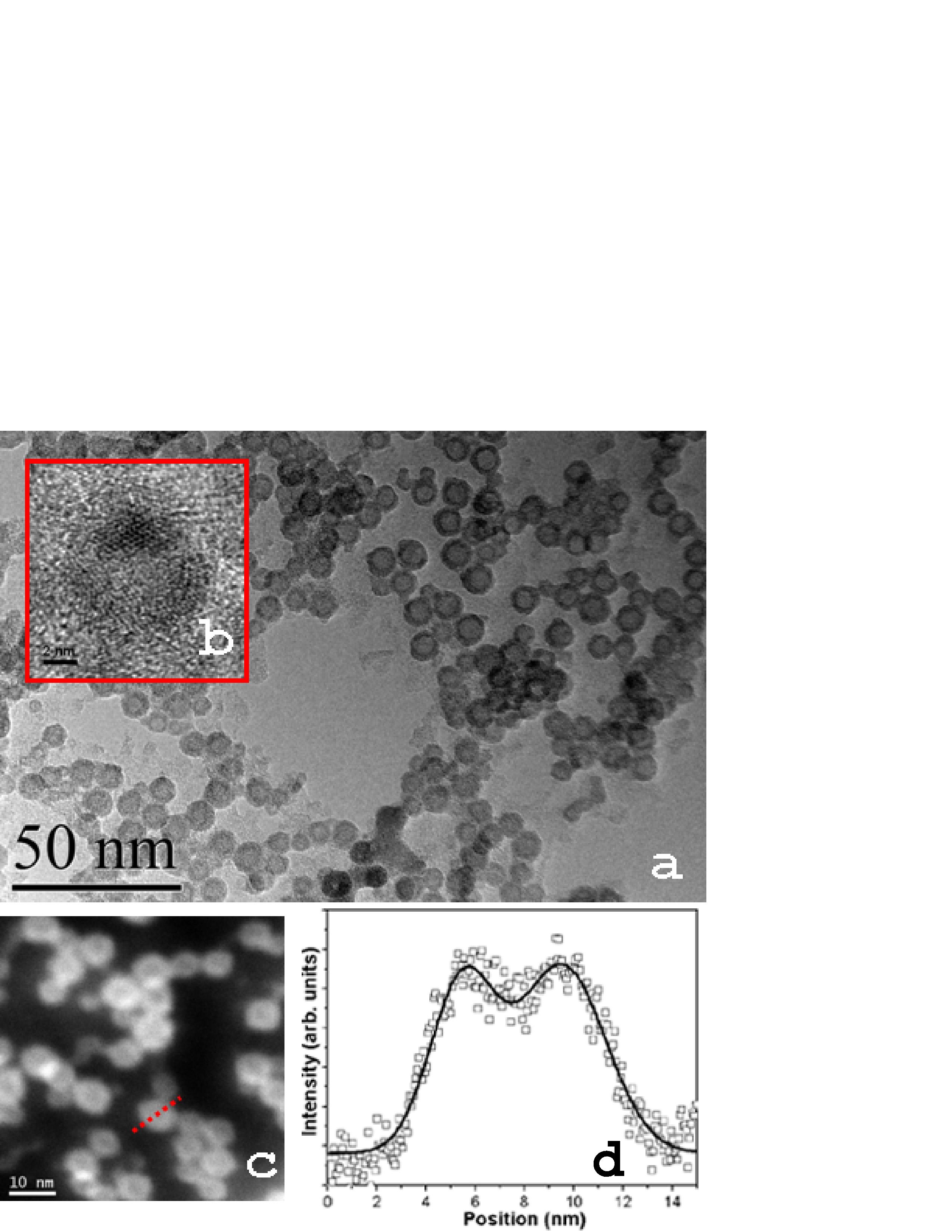}}
\caption{\label{FIG:TEM} a) TEM micrograph of the sample where the
projection of nanoparticles shows toroidal-like structures of 9.3 nm
in diameter, b) HRTEM obtained on one of the nanostructures c) HAADF
STEM images and d) STEM profile through one single nanoparticle
indicated in c) showing an enhanced contrast at the particle edge.}
\end{center}
\end{figure}
The morphology, structure and composition of the particles were
studied using High-resolution TEM (HRTEM) combined with Energy
Electron Loss Spectroscopy (EELS) and Energy Filtered TEM (EFTEM) as
well as high angular annular dark field (HAADF) and Bright Field TEM
(BFTEM). The samples were prepared by dropping a colloidal solution
of HNS onto a carbon-coated copper grid. Fig. \ref{FIG:TEM}-a shows
a general view of the sample in which 9.3 $nm$ in diameter
nanoparticles are observed. The iron oxide spinel structure obtained
from the Fourier transformation of the HRTEM (\ref{FIG:TEM}-b)
images is consistent with that obtained from X-ray diffraction.
Furthermore, broadening of the X-ray patterns reflects the existence
of crystallite sizes of 2 $nm$, much lower than the 9.3 $nm$
observed from HRTEM. Detailed analysis of magnified HRTEM revealed
the polycrystalline nature of the nanostructures with the absence of
any preferential orientation. It is important to point out that
depending on the defocus, crystal plains with d-spacing
corresponding to spinel structure were observed either on the
external part or on the top of the inner part. This discards the
existence of toroidal-like structures. EELS and EFTEM analysis (not
shown) have revealed that the only elements composing the
nanospheres were Fe and O. HAADF (Fig. \ref{FIG:TEM}-c) and STEM
(Fig. \ref{FIG:TEM}-d) analysis denotes a increase of the density at
the outer part. All these data reveal that the observed
nanoparticles are HNS. Moreover, BFTEM micrograph at high tilt angle
($35^{o}$) do not show changes on the observed morphology denoting
once more that our nanoparticles are HNS.
\begin{figure}
\begin{center}
\includegraphics[bb=-10bp 20bp 310bp 220bp,
clip,width=1.0\columnwidth,keepaspectratio]{{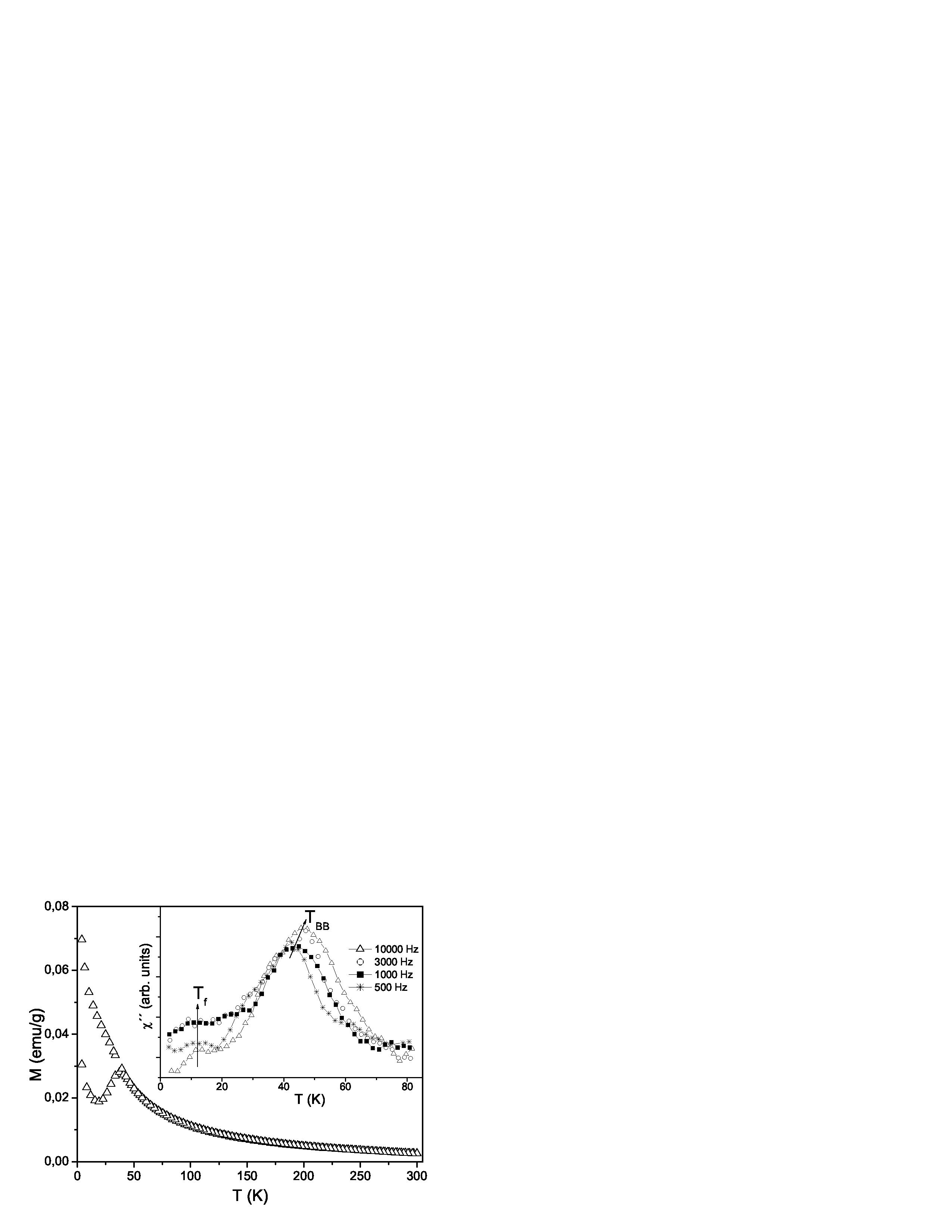}}
\caption{\label{FIG:ZFC} Magnetization as a function of temperature.
All curves are collected from 2 $K$ up to 300 $K$ with applied field
of $H=$ 20 $Oe$. ZFC (lower branch) and FC (upper branch) data are
shown. Inset: Thermal and frequency dependence of the out-of-phase
$\chi^{\prime\prime}$ component of the ac susceptibility.}
\end{center}
\end{figure}
\begin{figure}
\begin{center}
\includegraphics[bb=-20bp 185bp 375bp 415bp,
clip,width=1.0\columnwidth,keepaspectratio]{{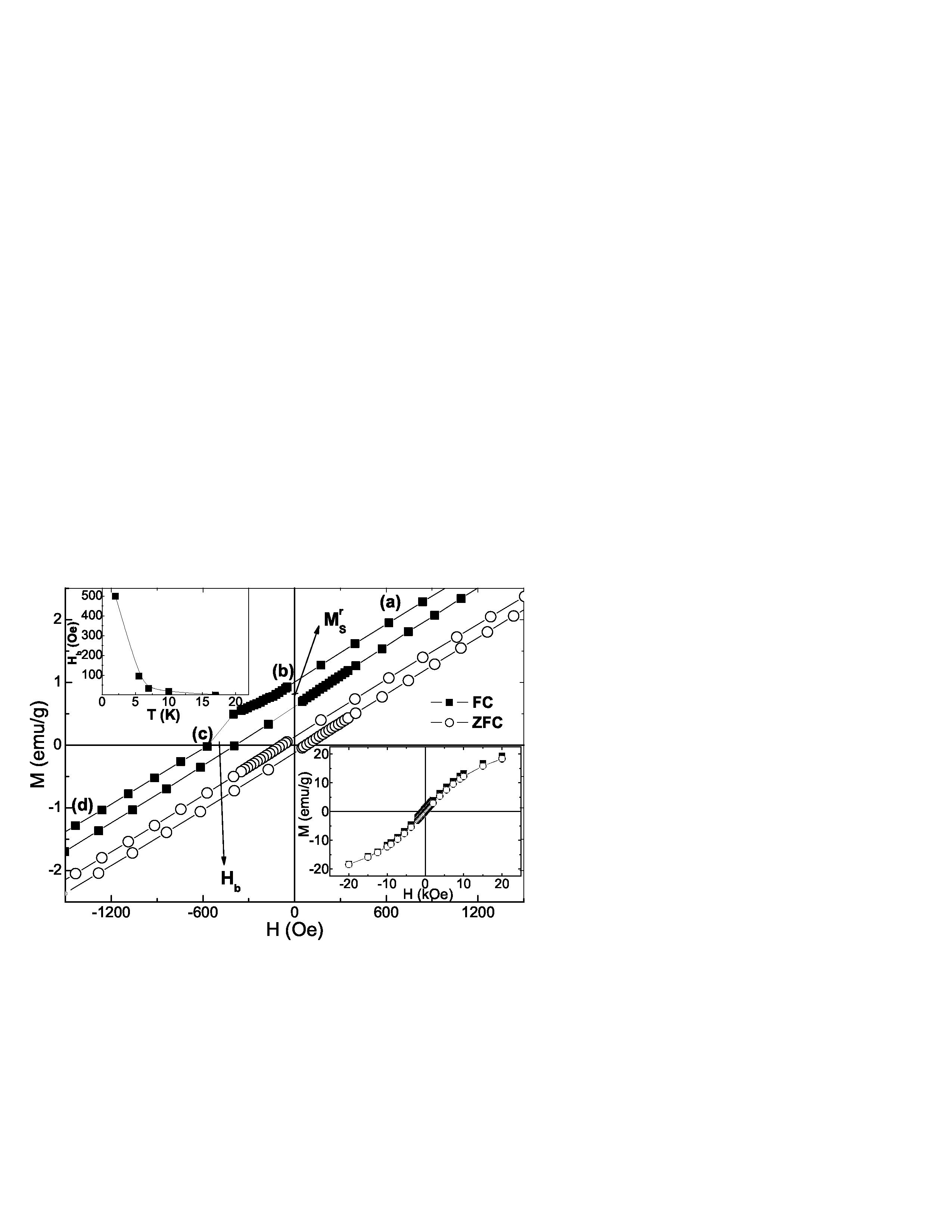}}
\caption{\label{FIG:MH} Low-field section of the magnetization
isotherms (in ZFC and FC modes) at $T =$ 2 $K$. Top inset: thermal
dependence of the bias $H_b(T)$ obtained from the FC $M(H)$ results.
Bottom inset: $M(H)$ measured up to 20 $kOe$. (a), (b), (c) and (d)
refer to Fig. \ref{FIG:DESENHO} (see text).}
\end{center}
\end{figure}
Figure \ref{FIG:ZFC} displays the $M(T)$ curves (H$=20$ $Oe$)
measured in the ZFC and FC (cooling field $H=$ 20 $Oe$). The ZFC
results exhibit a sharp peak at $\sim$ 36 $K$. This temperature
coincides with irreversibility temperature (i.e. the temperature
above which the ZFC and FC curves coincide). This is a indication of
the existence of a very narrow size distribution as observed in Fig.
\ref{FIG:TEM}-a. Unexpectedly, the ZFC magnetization as the FC
magnetization turns up below 20 $K$. This anomaly occurs at the same
temperature range where we observe the rise in $M_{FC}(T)$. The
inset of Fig. \ref{FIG:ZFC} displays the out-of-phase component
$\chi^{\prime\prime}$ of ac susceptibility as a function of
temperature under a magnetic field of 2 $Oe$ and at frequencies $f=$
0.5, 1, 3 and 10 $kHz$. The results exhibit two maxima located at
$T_{BB}=$ 45-55 $K$ and $T_f\sim$12 $K$. The maximum at $T_{BB}$ was
associated to the blocking process of the superparamagnetic (SPM)
magnetic moments in inner part of the HNS, hereafter (BULK).
$T_{BB}$ has a large dependence on frequency, which is an indication
of the existence of a thermally activated process, with an energy
barrier $E_a=1.8\times10^{-13}$ $erg$. Assuming that $E_a$ is the
product $K_{eff}V_{BULK}$ ($V_{BULK}$ is estimated from
M\"{o}ssbauer experiments as will be discussed later on), we obtain
$K_{eff} = 1.3 \times 10^6 erg/cm^3$ for this phase, slightly higher
than bulk magnetite. The second maximum at $T_f$ hardly change with
frequency and we associated it to the freezing of a cluster-glass
like phase (CGP) structure in the disordered uncompensated surface
regions (outer surface $S_1$ and inner surface $S_2$). The rise of
the ZFC and FC magnetization showed in Fig. \ref{FIG:ZFC} could be
associated to the uncompensated magnetic moment at the surface,
which will provide an increase of the ferrimagnetic moment below
$T_f$. Fig. \ref{FIG:MH} shows a detail of the magnetization loops
measured in the ZFC and FC (cooling field $H=$ 10 $kOe$). We
observed a large "loop shift" in FC cycle when measured at
temperatures below the CGP freezing temperature ($T_f <$ 20 $K$).
The bias field, $H_b$, is defined as the center field of the shifted
magnetic loop. Usually, in "core-shell" nanoparticles, $H_b$ is
associated to the bias anisotropy induced by the "exchange
interaction" between the magnetic microstructures in the frustrated
ordered shell pinned by a large surface magnetic anisotropy and the
soft ferromagnetically ordered core of the particle
\cite{MAR98,ZYS01}. In our case, we should understand the origin of
this bias field considering that the nanoparticles are not
"core-shell" but hollow nanospheres with two surface layers ($S_1$
and $S_2$) and the BULK inner region, as represented in figure
\ref{FIG:DESENHO}. In principle we can assume that $S_1$ and $S_2$
regions are highly frustrated magnetic layers of magnetic clusters
with a large surface magnetic anisotropy (responsible for the low
temperature freezing of the surface magnetic moments, $T_{f}$). The
inner BULK region shows a SPM behavior with low magnetic anisotropy.
When cooling the sample down to 2 $K$ under an external applied
magnetic field $H_0$, the resultant magnetization along the applied
field direction will be the sum of the contribution of the surface
regions ($\vec{M}_S$) and the BULK ($\vec{M}_B$):
$\vec{M}=\vec{M}_S+\vec{M}_B$ in the HNS (a). Once the field is
removed (b) at 2 $K$ (below $T_{f}$), the freeze magnetic moments
will keep a remanent magnetization ($\vec{M}_{S}^{r}$, see
hysteresis loop of Fig. \ref{FIG:MH}). We will consider that
$\vec{M}_S$ is contributed by $\vec{M}_{S}^{r}$, which is pinned
during the whole hysteresis cycle, and $\vec{M}_{S}^{up}$, which is
the unpinned field induced component at the surface
($\vec{M}_S=\vec{M}_{S}^{r}+\vec{M}_{S}^{up}$). When we apply an
external magnetic field opposite to the magnetization direction (c),
the magnetization within the BULK will rotate at low field values
(soft phase) but the freeze magnetic moments at $S_1$ and $S_2$ will
retain the magnetic state in which were frozen $M_{S}^{r}$. This
process will reduce $\vec{M}$ reaching $M=0$ at
$\vec{H_0}=-\vec{H_b}$, where $\vec{M_B}=-\vec{M_S}$. The field
necessary to compensate $M_{S}^{r}$ is the responsible of the "bias
field" (see Figure \ref{FIG:DESENHO}). Increasing $H_0$ an increase
of the magnetization is obtained favoring an alignment of the
magnetic moments along the $H_0$ in opposite direction to the (a)
and (b) situation. Increasing $H_0$ in the opposite direction of the
cooling field, we can reach the situation depicted in (d), which is
almost symmetric to (a). However, $\vec{M}_{S}^{r}$ contribution
remains and is the responsible for the shift of the hysteresis loop
toward positive values of magnetization when $H_0$ is again reduced.
$\vec{M}_{S}^{r}$ is originated in the FC process below $T_f$ and it
is absent when the sample is ZFC. Thus, $H_b$ disappears for
temperatures above $T_f$ or ZFC due to random alignment in the CGP
($M_{S}^{r}$).

\begin{figure}
\begin{center}
\includegraphics[bb=-40bp 121bp 290bp 300bp,
clip,width=1.0\columnwidth,keepaspectratio]{{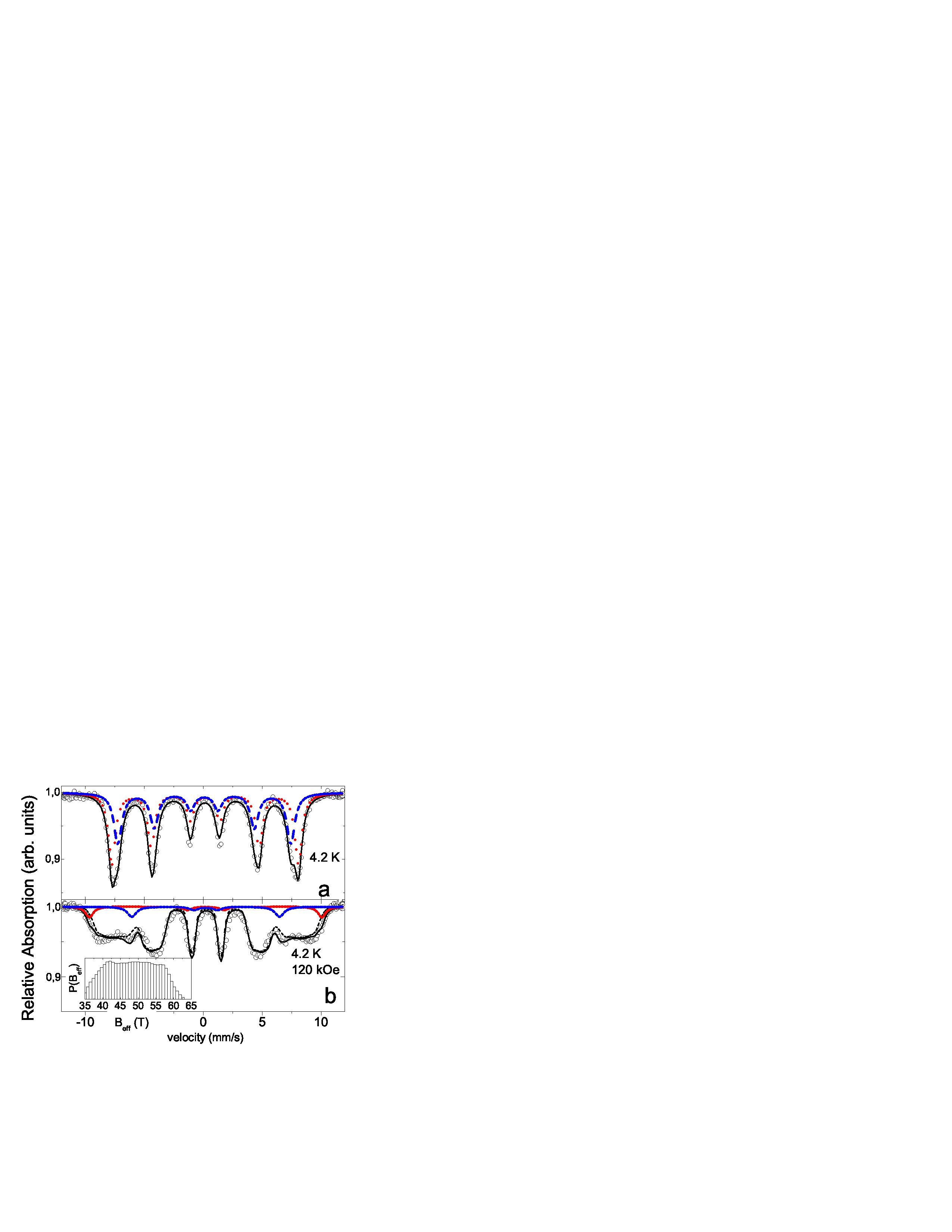}}
\caption{\label{FIG:MS} M\"{o}ssbauer spectra: a) Low temperature
4.2 $K$ and b) under applied field ($H_{app}=$ 120 $kOe$). Solid
line is the fitted spectrum and dashed red and blue lines are the
subspectra referent to sites A and B, respectively. Inset of figure
5-b is the hyperfine field distribution obtained from the fitting
procedure.}
\end{center}
\end{figure}

M\"{o}ssbauer spectra (MS) were taken at 4.2-300 $K$ in a liquid He
flow cryostat with a conventional constant-acceleration spectrometer
in transmission geometry using a $^{57}$Co/Rh source. For in-field
measurements, the powder sample was mounted in a vertical
source-sample-detector setup in the bore of a 140 $kOe$
superconducting magnet, such that the direction of $\gamma$-ray is
parallel to the applied field. The spectra were fitted by using
Lorentzian line shapes, and a foil of $\alpha-Fe$ was used to
calibrate the velocity scale. The room-temperature MS spectrum is a
doublet with narrow lines (line width $w =$0.65 $mm/s$), $IS$ =0.36
$mm/s$ and quadrupolar splitting $QS$ =0.98 $mm/s$. The $IS$ value
is similar to what is commonly observed in nanostructured ferrites
in SPM regime \cite{GOY03}. However, the $QS$ value, which
originates in the local charge density symmetry, is much larger than
the expected for these materials, reflecting a local symmetry lower
than cubic, which in turn will break the Fe-Fe superexchange paths
and/or oxygen vacancies located at both inner and outer surfaces.
This is consistent with the picture of a magnetically disturbed spin
configuration at the surface \cite{COE71}. At 4.2 K (Fig.
\ref{FIG:MS}-a) the relaxation time is slow enough to ensure a
static hyperfine splitting, and the spectrum could be fitted with
two sextets associated to sites A and B in the spinel-type
crystalline lattice. The obtained hyperfine field values ($B_{hf} =
491$ and 455 $kOe$ for sites A and B, respectively) are smaller than
bulk values \cite{ARE08}, an effect usually observed in core-shell
structures and assigned to the small $E_a$ (and the associated
softening) for the collective magnetic excitations which act to
reduce the hyperfine fields with respect to their values at $T =$ 0
$K$ \cite{MOR83}. However, in our case the $E_a$ value is of the
same order than observed for crystalline magnetite nanoparticles.
Thus, the origin of reduced $B_{hf}$ is the surface disorder. The
large linewidth values of the magnetic sextets (Fig. \ref{FIG:MS}-a)
also indicate a locally disordered environment of Fe ions. In MS
experiments under applied field, the effective hyperfine field
$B_{eff}$ will be the vector sum of the applied field $H_{app}$ and
the hyperfine field $B_{hf}$. Because of the strong
antiferromagnetic interaction between sublattices A and B in the
ferrites, $B_{eff}$ of the sub-lattice A increases while $B_{eff}$
of the sub-lattice B decreases. Moreover, the relative intensities
of the six-line MS spectra are given by: $3:p:1:1:p:3$,
$p=4\sin^{2}\alpha/(1+\cos^{2}\alpha)$, where $\alpha$ is the angle
between the spin and the gamma-ray direction \cite{DIC86}.
Therefore, lines 2 and 5 vanish when the magnetic moments of the
particles align to the applied field. Fig. \ref{FIG:MS}-b shows the
MS spectra measured at 4.2 K under $H_{app} =$ 120 $kOe$, which is
composed of very broad, strongly overlapping lines. Considering our
TEM and magnetic data, we proposed a fitting procedure based on the
combination of two crystalline sextets with narrow lines plus a
continuous $P(B_{eff})$ distribution consistent with the existence
of the (CGP). The crystalline sextets were assumed to correspond to
spins in A and B sites aligned with applied field (red and blue
subspectra, respectively). In addition, the intensities of lines 2
and 5 are fixed at 0 for these components. The relative area of
these crystalline subspectra is 6-7 $\%$, showing that only a small
fraction of spins, probably located in the BULK region, are aligned
to the external field. The hyperfine field distribution resulting
from the fitting is displayed in the inset of Fig. \ref{FIG:MS}-b,
and it is associated to the sites not aligned with $H_{app}$. Its
contribution amounts to 87 $\%$, reflecting a high fraction of
misaligned moments. If we consider that the surface moments are the
only one disordered at high field, we can estimate a thickness of
0.9 $nm$ for $S_1$ and $S_2$ regions and an inner diameter of 4 $nm$
(empty region) compatible with STEM profile (see Fig.
\ref{FIG:TEM}-d). The effective field distribution is very broad,
with an equivalent probability for all $B_{eff}$ values between
those obtained for the two crystalline sextets. In addition, the
ratio between the line 2 and 3 ($I_{23}$) for this component is very
close to 2, the same as for a randomly oriented sample. This result
supports the picture of a spatially disordered freezing of a large
amount (87$\%$) of spins residing in the morphologically disordered
surface areas.

Summarizing, we propose a new synthesis method to obtain ferrite
hollow nanospheres. The magnetic characterization of this type of
nanoestructures brings about relevant phenomena which are explained
within a simple model based on the coexistence of a SPM soft phase
(BULK) and the CGP hard phase ($S_1$ and $S_2$).

\begin{acknowledgments}
We acknowledge the critical reading and discussion with Prof. P. A.
Algarabel. This work received financial supported from Argentinian
ANPCyT, CONICET and UNCuyo, Spanish MEC under grants:
NAN2006-26646-E and Consolider CSD2006-12, and Brazilian FAPESP and CNPq. G.
F. G. acknowledges support from the Spanish MEC through the Ramon
y Cajal program. E. L. Jr. acknowledges fellowship by FAPESP.
\end{acknowledgments}

\end{document}